\documentstyle[aps,psfig,epsfig,amssymb,bbm]{revtex}

\newcommand{\Hb}{{\hat{\mathcal H}}_{\beta}}

\newcommand{\geqsi}{\; {\scriptstyle {> \atop \sim}} \;}

\newcommand{\ep}{\epsilon}
\newcommand{\ps}{\ps}

\newcommand{\Ih}{{\hat I}}

\newcommand{\Up}{{\hat{\mathcal U}}}

\newcommand{\Np}{{\hat{\mathcal N}}}

\newcommand{\Ub}{\Up_{\beta}}

\newcommand{\epsm}{$\epsilon$SM}

\begin{document} 
 
\draft
\title{A classical scaling theory of quantum resonances}
 
\author{Sandro Wimberger,$^{1,2,3}$ Italo~Guarneri,$^{2,3,4}$ and  
Shmuel~Fishman$^{5}$}  
\address{$^{1}$Max-Planck-Institut f\"ur Physik komplexer Systeme, 
N\"othnitzer Str. 38, D-01187 Dresden \\ 
$^{2}$Universit\`a degli Studi dell' Insubria, Via Valleggio 11, 
I-22100 Como\\ 
$^{3}$Istituto Nazionale per la Fisica della Materia, Unit\`a di Milano, 
Via Celoria, I-20133 Milano \\ 
$^{4}$Istituto Nazionale di Fisica Nucleare, Sezione di Pavia, Via Bassi 6, 
I-27100 Pavia\\ 
$^{5}$Physics Department, Technion, Haifa IL-32000 \\ 
} 
\date{\today}
\narrowtext
\twocolumn
\maketitle 

\begin{abstract} 
The quantum resonances occurring with $\delta-$kicked particles 
 are studied with the help 
of a fictitious classical limit, establishing a direct correspondence 
between the nearly resonant quantum motion and the classical 
resonances of a related system. A scaling law 
which characterizes the structure of the resonant peaks is 
derived, and numerically demonstrated. 

\end{abstract} 
 

Resonances are a widespread phenomenon both in quantum and in 
classical physics. However, quantum resonances are often purely quantal 
phenomena, not directly related to classical nonlinear resonances. 
Atom optics has made 
crucial aspects of the classical-quantum 
correspondence accessible to direct experimental observation. 
An outstanding example is the paradigmatic Kicked Rotor (KR) Model, which 
served as a prototype 
of quantum and classical Hamiltonian chaos over  
almost three decades. This model 
has been  experimentally  realized  by a technique first introduced 
by Raizen and co-workers \cite{raizen}, 
allowing for observation of 
its
main physical properties. 
These are connected 
with the long-time evolution, and depend on the arithmetic 
nature of the effective Planck's constant (which is the Planck constant 
divided by a characteristic action) \cite{haake2001}. 
If the latter is sufficiently 
irrational, the rotor's energy saturates in time. This effect is known 
as dynamical localization. If 
the effective Planck's constant is a rational multiple of 
$4\pi$ ($=4\pi r/q$, with $r,q$ integers) 
unbounded growth of energy typically occurs. These are known 
as quantum resonances and have no counterpart in the 
corresponding classical system. 
Besides confirming theoretically known facts, 
experiments have also introduced variants of the basic model, 
which have enriched the theory. 
A striking example was  the experimental discovery of unexpected 
 ``quantum accelerator modes'' due to effects of gravity 
\cite{schlunk2003}.

Even in cases when  no major variants are introduced,  
experiments inevitably involve deviations from the 
idealized  theoretical model. The most obvious of these 
is that atoms move in lines and not in circles; though to some extent 
trivial, this fact  imposes certain modifications on the standard 
KR theory, especially in the case of the quantum resonances \cite{FGR2002}.   
Another experimental limitation is that experiments 
are confined to not too long  times. Consequently, they can neither detect the 
high-order (high$-q$) quantum resonances, nor address  
the basic, as yet 
unsolved, mathematical problem of what degree of irrationality is 
required  for  localization. 
Nevertheless, for this reason they have 
encouraged  fruitful theoretical analysis of the 
intermediate time regime, 
which had rarely been investigated in fine detail. 

In this letter we present  experimentally relevant  
results concerning the manifestations of quantum 
resonances at finite observation times. Our analysis enlightens the 
connections between quantum and classical resonances.   
 When the energy of the kicked atoms is measured after a fixed time, 
and the result is plotted versus the kicking period, resonances 
manifest in the form of peaks at the resonant values of the 
period (see Fig.1).  We describe  the line shape of such 
resonant peaks, and its dependence on the observation time and on 
the other relevant  parameters. Our main result is a single-parameter 
scaling law for such peaks, which  in particular implies an inverse 
square dependence of the peak width on time, and
is probably related to the ``sub-Fourier resonances'' recently
detected  in a different atom-optical context \cite{szriftgiser2002}. 
Our derivation is based on a  quasi-classical analysis of   
the purely quantal resonances and their vicinity. This seemingly 
self-contradictory task is accomplished by 
establishing a direct correspondence between the 
quantum resonances and the classical resonances of a related 
system, which is {\em not} obtained in the usual classical limit of 
vanishing Planck's constant, but rather in the limit when the 
effective Planck's constant approaches 
its resonant value, which is of order unity. In this limit, the 
detuning of the driving period from resonance plays the role of the 
Planck constant. The same technique was used in \cite{FGR2002} 
to explain the quantum accelerator 
modes observed in the presence of gravity. Here it is applied in 
the gravity-free case, leading to a different physical scenario. 
It may also be 
adapted to the important case when decoherence is added \cite{WGF2003}. 

The specific system we consider is described, in 
dimensionless units, by the Hamiltonian 
\begin{equation}
{\cal H }(t') =\frac{\tau}{2} p^2 + k\cos(x)\sum_{t=-\infty}^{+\infty}
 \delta (t'-t)\;,
\label{ham}
\end{equation}
where $x$ is the coordinate, $p$ is its conjugate momentum, 
$t'$ is the continuous time variable, while $t$ is an 
integer that counts the number of kicks. If $x$ is an angle 
variable, confined to the 
interval $[0, 2 \pi)$, then 
(\ref{ham})  is the Hamiltonian of the KR and 
$p$ is the angular momentum; when quantized, it takes integer 
values only. The two dimensionless
 parameters that control the dynamics are the effective 
Planck's constant $\tau$ and the strength of the kick $k$. 
The first (second) is proportional (inversely proportional)
to Planck's constant  $\hbar$, therefore $k \tau = K$ is 
independent of $\hbar$. $K$ is the  
stochasticity parameter of the Standard Map, which 
rules the classical dynamics defined by (\ref{ham}). 
The energy levels $\tau n^2/2$  ($n$ integer) 
of the free rotor ($k=0$), and  differences between them,  
are integer multiples of $\tau/2$, 
while the frequencies of the driving 
potential are integer multiples of $2\pi$, 
consequently quantum resonances of the kicked rotor 
(\ref{ham}) are found if  $\tau/4\pi$ is rational 
\cite{izrailev79}. These resonances 
correspond to the Talbot effect in optics and  
$4\pi$ corresponds to the Talbot time \cite{DB1997}. 
For most rational $\tau/4\pi$ they  result in ballistic growth 
of $p$ (quadratic growth of energy) with time. 
\begin{figure}[t]
\centerline{\epsfig{figure=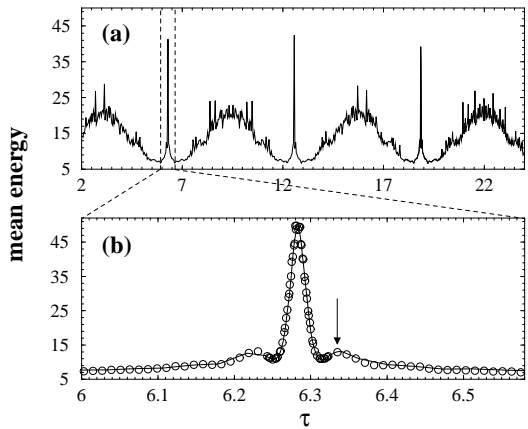,width=5.8cm,angle=270}} 
\caption{(a) mean energy after $t=30$ kicks, vs. 
the kicking period $\tau$, for an  ensemble of 
$10^5$ $\delta-$kicked atoms, with Gaussian initial momentum distribution 
($\sigma \simeq 2.7$) and $k=0.8\pi$. (b) 
quantum data are compared with the mean energies of an ensemble of $10^6$ 
particles (circles), evolving under the $\ep-$classical dynamics 
(\ref{clmap}). The 
small peak on the right of the resonant spike is marked by an arrow 
for reference to Fig.~\ref{fig2}. 
(The peaks appear lower in (a) because the used computational grid in $\tau$ 
had a lower resolution than in (b).)
}
\label{fig1} 
\end{figure}
If $x$ is not a coordinate on a circle but on the full line 
instead, then $p$ is linear momentum 
and (\ref{ham}) yields  the Kicked Particle (KP) 
model which is the
one relevant for recent atom optics experiments. 
Since the driving potential is periodic in $x$ 
only transitions between values of $p$ that differ by an integer 
are allowed. Consequently, the quasimomentum (QM)  
$\beta$ (the fractional part of  $p$) is 
conserved \cite{FGR2002,WGF2003}. 
Standard Bloch theory allows for reduction of  the 
KP dynamics to the dynamics of a bundle of rotors. 
At any given (and constant in time) 
value of $\beta$, the dynamics induced by  (\ref{ham}) is formally 
that of a rotor  and the one-period Floquet operator for this rotor 
is  \cite{FGR2002}: 
\begin{equation} 
\label{ubeta} 
\Ub \;=\;e^{-{\rm i} k\cos({\hat \theta})}\;e^{-{\rm i}\tau(\Np+\beta)^2/2}, 
\end{equation} 
where $\theta=x$mod$(2\pi)$, and 
$\Np=-{\rm i}d/d\theta$ is the angular momentum operator. 
If $\tau/4\pi$ is rational, the KP dynamics may exhibit 
asymptotic quadratic growth of the energy only for 
finitely many special values of the QM $\beta$. The  
particle is then in a spatially  extended  state \cite{WGF2003}. 
Wave-packet dynamics, or incoherent ensemble averaging over momentum  
as is the case in experimental realizations \cite{darcy2001}  
enforces {\it linear} growth of the energy as a function of time for rational 
$\tau/4\pi$ \cite{FGR2002,WGF2003}. Away from resonances, the 
growth of energy in time is bounded owing to dynamical 
localization \cite{haake2001}, so 
a scan of the particle energy vs. $\tau$ 
after a fixed evolution time exhibits peaks at the resonant values of $\tau$, 
as shown in Fig. 1. Such peaks rise and shrink as the observation time 
is increased. Their height is proportional to time, and their width 
to inverse {\it square} time, as our present analysis will show. \\
We restrict  to the main resonances 
and to their vicinity, that is to $\tau=2\pi \ell +\ep$ where 
$\ell$ is a positive 
integer and $\ep$ is small compared to $2\pi$. 
We denote $\tilde k=k|\epsilon|$, and 
$\Ih=|\epsilon|\Np$. Then, using the identity 
$e^{{\rm i}\pi\ell n^2}=e^{{\rm i}\pi\ell n}$,  
eq. (\ref{ubeta}) may be rewritten as \cite{FGR2002,WGF2003}
\begin{equation} 
\label{onecycleeps} 
 \Ub (t)\;=\;e^{-{\rm i}  
\tilde k\cos({\hat \theta})/|\ep |}\;e^{- {\rm i}\Hb/|\ep |}\;,
\end{equation} 
where 
\begin{eqnarray} 
\label{epsscal} 
\Hb (\Ih,t) &=& \frac{1}{2} 
\mbox{sign}(\ep) \Ih^2 + \Ih (\pi \ell +\tau \beta)\;.
\end{eqnarray} 
If $|\epsilon|$ is regarded as Planck's constant \cite{casati},
then (\ref{onecycleeps}) is the formal quantization 
of either of the  following classical maps: 
\begin{eqnarray} 
\label{clmap} 
I_{t+1} &=&I_{t}+{\tilde k}\sin (\theta_{t+1})\;, \nonumber \\
\theta_{t+1}&=&\theta_t\pm I_t+\pi \ell+\tau \beta\;\;\mbox{\rm mod} 
(2\pi)\, 
\end{eqnarray} 
where $\pm$ is the sign of $\epsilon$. 
The small$-|\epsilon|$ asymptotics of the quantum 
$\beta-$rotor is thus equivalent to a quasi-classical approximation    
based on the $\ep-$classical dynamics  (\ref{clmap}). 
Here the term  ``classical'' is not 
related to the $\hbar\to 0$ limit but to the limit $\epsilon\to 0$ instead,  
so the term ``$\ep-$classical'' will be used in the following.
 Note that the effective Planck's constant 
$\tau$  {\em is not small}, and important quantum 
symmetries are taken into account in the $\ep$-classical 
description. 
The efficiency of this approximation is demonstrated in Fig.1(b), where 
results of exact quantum simulations are compared to results obtained 
by (\ref{clmap}), for an 
initial incoherent ensemble of particles, each in a momentum eigenstate
(a plane wave) 
and with a Gaussian distribution of the momentum $p=n_0+\beta$.  
For each particle in the ensemble, the map (\ref{clmap})  
with $\beta$ equal to the QM of the particle was used to compute  
a set of trajectories started at  
$I=n_0 |\ep|$ with homogeneously distributed $\theta\in[0,2\pi)$.  
The final energies $\ep^{-2} I_t^2/2$ at 
$t=30$  of the individual trajectories were averaged over $\theta, 
\beta,n_0$ with the weights imposed by the initial ensemble.  
The average energy $\langle E_t\rangle=\ep^{-2}\langle I_t^2\rangle/2$ 
is plotted vs. $\tau=2\pi+\ep$ in Fig.~\ref{fig1}(b), 
along with results of quantal computations of the ensemble averaged energy.\\ 
We shall show that the resonant peaks 
are determined by $\ep-$classical phase space structures.
It is convenient to change variables to $J=\pm I+\pi\ell+\tau\beta$, 
${\vartheta}=\theta+\pi(1-\mbox{\rm sgn}(\ep))/2$, thus  turning the maps    
(\ref{clmap}) into a single $\epsilon$-classical 
Standard Map (\epsm ), independent of the value of $\beta$:
\begin{equation}
\label{epsmap}
J_{t+1}=J_t+{\tilde k}\sin (\vartheta_{t+1})\;\;,\;\;
\vartheta_{t+1}=\vartheta_t+J_t\;.
\end{equation}
It will turn out that: 
(i) The resonant value $\tau=2\pi\ell$ corresponds to integrable 
$\ep-$classical dynamics, and the resonant values of the QM  
$\beta$ correspond to the $\ep-$classical resonant values of the  
action $J$; (ii) for small $|\ep|$, 
the $\ep-$classical dynamics is quasi-integrable, 
and the growth of the quantum energy is dominated by the 
main $\ep-$classical resonant island around $J=2\pi$;
(iii) at any time $t$, the ratio between the energy and its value at 
$\ep=0$ is a scaling function, notably it is a function of the single 
variable $x=t\sqrt{k|\ep|}$ and not of the variables $t,k,\ep$ 
separately.\\
We assume for simplicity an initially flat distribution of 
$p\in [0,1]$; then $I_0=0$, and $J_0=\pi\ell+\tau\beta_0$ 
with $\beta_0$ uniformly distributed in $[0,1)$. Without loss of generality 
we also consider $\ell=1$. Hence if $|\ep|\ll1$ 
then $J_0$ is uniformly distributed over one period (in action) 
$(\pi, 3\pi)$ of the \epsm.  
The resonant QM value is  $\beta=1/2$ \cite{WGF2003}. 
Since $J_t=\pm I_t+\pi+\tau\beta$, and $I_0=0$, the  
mean energy of the rotor at time $t$  is: 
$$ 
\langle E_{t,\ep}\rangle=\ep^{-2}\langle I_t^2\rangle/2=
\ep^{-2}\langle(\delta J_t)^2\rangle/2\;,\;\;
\delta J_t=J_t-J_0\;. 
$$
The exact quantum resonance  $\ep=0$ corresponds to the 
integrable limit of the \epsm, where $\delta J_t=0$. However, 
$\langle E_{t,\ep}\rangle$ is scaled by $\ep^{-2}$, so in order 
to compute it  at $\ep=0$ one has 
to compute $\delta J_t$ at first order in $\ep$. This leads to:
\begin{equation}
\label{Igrth}
\delta J_t\;=\;|\ep|k\sum\limits_{s=0}^{t-1}\sin(\theta_0+J_0 s)+
r(\ep,t)
\end{equation}
where $r(\ep,t)=o(\ep)$ as $\ep\to 0$ at fixed $t$. 
The particle energy at time $t$ is found from (\ref{Igrth}) by 
taking squares,  averaging over $\theta_0$, $J_0$,  
dividing by $2|\ep|^2$, and finally letting $\ep\to 0$:
\begin{equation}
\label{resgrth}
\langle E_{t,0}\rangle=\frac{k^2}{8\pi}\int_{\pi}^{3\pi}
dJ_0\;\frac{\sin^2(J_0t/2)}{\sin^2(J_0/2)}=\frac{k^2}{4}t\;.
\end{equation}
The small contribution of the initial QM  
in the atom's energy was neglected. Apart from that, 
(\ref{resgrth}) is equal to the exact quantum mechanical result \cite{WGF2003}.
The integral over $J_0$ in (\ref{resgrth}) collects 
contributions from all the invariant tori $J_0=$const. of the 
\epsm~ at $\ep=0$, but it is essentially determined by 
a small interval $\sim 2\pi/t$ of 
actions around $J_0=2\pi$, the main $\ep-$classical resonant torus.
As that torus is  formed of (period 1) fixed points, its own  
contribution is quadratic in time, so  the linear 
growth (\ref{resgrth}) follows.
Noting  that $J_0=2\pi$ corresponds to 
the resonant QM: $\beta_0=1/2$, we see that 
the $\ep-$quasi-classical approximation explains  the {\it quantum} 
resonances in terms of  {\it classical} resonances of the 
Standard Map \cite{LL92}. 
We now estimate  $\langle E_{t,\ep}\rangle$ for $|\ep|>0$. 
The $|\ep|>0$ dynamics 
is maximally distorted with respect to the $\ep=0$ one for $J_0$ in 
the vicinity of  $2n\pi$, that is, in the very region which is mostly 
responsible for the linear growth of energy at $\ep=0$.  
Being formed of period-1 fixed points,  
the $J_0=2\pi$, $\ep=0$  invariant torus breaks  at $|\ep |>0$ 
as described by the Poincar\'e-Birkhoff  
theorem \cite{LL92}, giving rise 
to the ``main resonance'' of the \epsm, which is located astride $J=2\pi$ 
with a size $\delta J_{res}\approx 4(k|\ep|)^{1/2}$ \cite{LL92}.  
The approximation 
(\ref{Igrth}) fails quickly therein so its contribution 
$\langle E_{t,\ep}\rangle_{res}$ in the mean energy 
has to be estimated differently. 
In the remaining part of the $\ep-$classical 
phase space the motion mostly  follows KAM invariant curves, 
slightly deformed with respect to the $\ep=0$ ones, still with  
the same rotation angles. The contribution of 
such invariant curves  
in the mean energy is therefore roughly similar to that 
considered in the integral (\ref{resgrth}).
 On such grounds, in order to estimate 
$\langle E_{t,\ep}\rangle$ we remove from the integral (\ref{resgrth}) 
the contribution of the resonant action interval near $J_0=2\pi$, and we 
replace it by: 
\begin{equation}
\label {sup}
\langle E_{t,\ep}\rangle\sim\frac{k^2}{4}t-\Phi(t)+\langle 
E_{t,\ep}\rangle_{res}\;,
\end{equation}
where
\begin{equation}
\label{phires}
\Phi(t)=\frac{k^2}{8\pi}\int_{-\delta J_{res}/2}^{\delta J_{res}/2}
dJ'\;\frac{\sin^2(tJ'/2)}{\sin^2(J'/2)}\;,
\end{equation}
and $J'$  is the deviation from the resonant value $2 \pi$. 
$\langle E_{t,\ep}\rangle_{res}$ 
may  be estimated  by means of the 
pendulum approximation \cite{LL92}. Near the \epsm~ 
resonance, the motion is described (in {\it continuous} time) 
by the pendulum Hamiltonian in the coordinates $J',\vartheta$:
$H_{res}=\frac12 (J')^2+|\ep|k\cos(\vartheta)$.
A characteristic time scale for the motion in the resonant zone is 
$t_{res}=(k|\ep|)^{-1/2}$, which is the period of the small pendulum 
oscillations divided by $2\pi$.
One may altogether remove $|\ep|$ from the Hamilton equations, 
by scaling momentum and time by factors $(k|\ep|)^{-1/2}=
4/\delta J_{res}$, 
$(k|\ep|)^{1/2}=1/t_{res}$ respectively. Therefore, 
\begin{equation}
\label{enspend}
\langle(\delta J_t)^2\rangle=\langle(J'_t-J'_0)^2\rangle\sim 
k|\ep| G(t\sqrt{k|\ep|})\;,
\end{equation}
for an  ensemble of orbits started inside the resonant zone,  
where $G(.)$ is a parameter-free function. 
This function results of averaging over 
nonlinear  pendulum motions with a continuum of different periods, 
so it saturates  to a constant  value when 
the argument is larger than $\approx 1$.    
The contribution to the total energy is then obtained on 
multiplying (\ref{enspend}) by $|\ep|^{-2}\delta J_{res}/(4\pi)$, 
because only a fraction $\sim \delta J_{res}/(2\pi)$ of the initial 
ensemble is  trapped in the 
resonant zone. As a result 
\begin{equation}
\label{enspend1}
\langle E_{t,\epsilon}\rangle_{res}\sim \pi^{-1}|\ep|^{-1/2}k^{3/2}
G(t\sqrt{k|\ep|})\;.
\end{equation}
When $\delta J_{res}$ is small, $\sin^2(J'/2)$ may be replaced by $J'^2/4$ 
in the integrand in (\ref{phires}), leading to 
$$
\Phi(t)\sim \frac{k^2}{4}t\;\Phi_0(t\sqrt{k|\ep|})\;,
\Phi_0(x)\equiv\frac{2}{\pi}\int_0^x
ds\;\frac{\sin^2(s)}{s^2}\;.
$$
\begin{figure}[t]
\centerline{\epsfig{figure=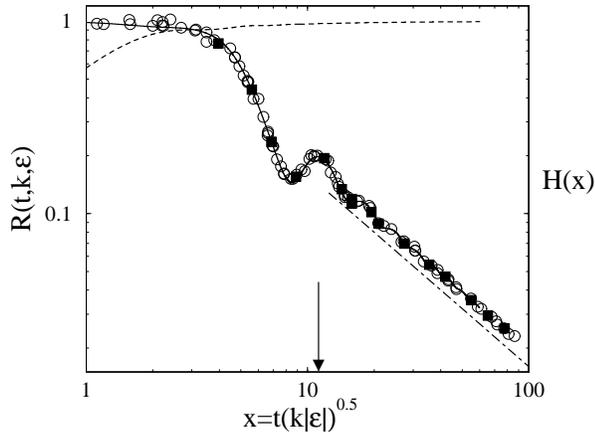,width=6cm,angle=270}} 
\caption{Demonstrating the scaling (\ref{repla0}) at $\tau \geqsi
2\pi$. Open circles correspond 
to different values of the parameters $\ep,k,t$, randomly and independently 
generated  in the ranges $1<t<200$, $0.001<\ep <0.1, 
0.1<k<50$ with the constraint $0.01<k\ep <0.2$.
In each case an ensemble  
of $2\times 10^6$ $\ep-$classical rotors was used to numerically 
compute the scaled energy $R(t,k,\ep)$ (\ref{repla0}), with a 
uniform distributions of initial 
momenta in $[0,1]$ and of initial $\theta$ in $[0,2\pi)$. 
Full squares present quantum data for $k=0.8 \pi, t=50$ and $t=200$. 
The solid line 
is the scaling function $H(x)$ of (\ref{repla0}) obtained by 
computation of the functions  $\Phi_0(x)$ (dashed) and $G(x)$. 
The dot-dashed line has slope $-1$. The arrow marks the value of the scaled 
detuning $x$ corresponding to the arrow in Fig.1(b). 
}
\label{fig2}
\end{figure}
Replacing in (\ref{sup}), we obtain:
\begin{eqnarray}
\label{repla0}
R(t,k,\ep)&\equiv& \frac{\langle E_{t,\ep}\rangle}
{\langle E_{t,0}\rangle}\sim H(x)\equiv 
1-\Phi_0(x)+\frac{4}{\pi x}G(x)\;,\nonumber\\
&x&=t\sqrt{k|\ep|}
=t/t_{res}\;.
\end{eqnarray}
Hence  $R(t,k,\ep)$ depends on $t,k,\ep$ only through 
the scaling variable 
$x=t/t_{res}$.  
The width in $\ep$ of the resonant peak therefore scales like 
$(kt^2)^{-1}$. The scaling law (\ref{repla0}) is demonstrated  
by numerical data shown in Fig.~\ref{fig2}. 
The function $H(x)$ was numerically computed: in particular, 
$G(x)$ was obtained by a standard Runge-Kutta integration  of the 
pendulum dynamics.  The scaling function $H(x)$ 
decays proportional to $x^{-1}$ at large $x$, because so 
do  $1-\Phi_0$ and $4G(x)/(\pi x)$; the latter, owing to the 
saturation of $G$. Since $\Phi_0$ 
is quite slowly varying  at $x> 4$, the  
structures observed in that region are due to $G(x)$, which  
describes the resonant island.   \\
Our  analysis neglects higher-harmonics  
resonances of the \epsm,  higher order islands, and especially 
the growth of the stochastic layer surrounding the main 
resonance \cite{WGF2003}. It is therefore valid only if 
$k|\epsilon|$ is much smaller than $1$, which is roughly the threshold 
for global chaos. In the case when the smooth initial momentum 
distribution 
 includes values $n_0\neq 0$ and/or is appreciably non-uniform 
in QM, the statistical weights of the various phase-space 
regions are different. Scaling in the single variable $t/t_{res}$ still 
holds, but the scaling functions $\Phi_0$ and $G$ may be different. \\
The map (\ref{clmap}) is easily 
adapted to  the model in the presence of  decoherence due to 
random momentum jumps induced by external noise, e.g. the 
Spontaneous Emissions effects used in experiments 
\cite{darcy2001,klappauf98}. A scaling law is again valid for 
the resonant peaks, in the {\it two} variables 
$t/t_{res}$ and $t_c/t_{res}$  where $t_c$ is the time 
scale associated with noise \cite{WGF2003}.  \\
To summarize, we have exploited  a correspondence between 
the dynamics of a Kicked Particle 
near the quantum resonances  $\tau=2\pi \ell$ and the classical dynamics 
of a quasi-integrable system 
to analyze the structure of the experimentally observable 
quantum resonant peaks. For these we have derived a scaling law  
where the scaling variable is
the ratio between the observation time and 
the characteristic time scale of elliptic motion inside the $\epsilon-$
classical island. This law provides significant new information 
about the shape and the parameter dependence of the peaks.

We acknowledge discussions with A. Buchleitner, M. d'Arcy, S. Gardiner, 
R. Godun, M. Oberthaler, and G. Summy, and support by the INFM-PA project 
{\it Weak Chaos: theory and applications}, the EU 
QTRANS RTN1-1999-08400, the US-Israel BSF, 
the Minerva Center of Nonlinear Physics of Complex Systems,
and the fund for Promotion of Research at the Technion.

\end{document}